\documentclass[10pt,letterpaper]{article}
\usepackage{opex3}
\usepackage{setspace}
\usepackage{amssymb}

\begin{document}
%
%

\title{Spatial and temporal coherence properties of single free-electron laser pulses}

\author{A. Singer$^1$, F. Sorgenfrei$^2$, A. P. Mancuso$^3$, N. Gerasimova$^1$, O. M. Yefanov$^1$, J. Gulden$^1$,
T. Gorniak$^{4,5}$, T. Senkbeil$^{4,5}$, A. Sakdinawat$^{6}$, Y. Liu$^7$, D. Attwood$^7$, S. Dziarzhytski$^1$,
D. D. Mai$^8$,
R. Treusch$^1$, E. Weckert$^1$, T. Salditt$^8$,
A. Rosenhahn$^{4,5}$, W. Wurth$^{2*}$ and I. A. Vartanyants$^{1,9*}$}

\address
{$^1$ Deutsches Elektronen-Synchrotron DESY, Notkestr. 85, D-22607 Hamburg, Germany\\
$^2$Institut f\"ur Experimentalphysik and CFEL, University of Hamburg, Luruper Chaussee 149, 22761 Hamburg, Germany\\
$^3$European XFEL GmbH, Albert-Einstein-Ring 19, 22761 Hamburg, Germany\\
$^4$University of Heidelberg, Im Neuenheimer Feld 253, 69120 Heidelberg, Germany\\
$^5$Institute of Functional Interfaces, Karlsruhe Institute of Technology, Hermann-von-Helmholtz-Platz 1, 76344 Eggenstein-Leopoldshafen, Germany\\
$^6$SLAC National Accelerator Laboratory, 2575 Sand Hill Road, Menlo Park, California 94025-7015, USA\\
$^7$University of California, Berkeley, CA 94720, USA\\
$^8$Institut f\"ur R\"ontgenphysik, Georg-August-Universit\"at G\"ottingen, Friedrich-Hund-Platz 1,
D-37077 G\"ottingen, Germany\\
$^9$National Research Nuclear University, ''MEPhI'', 115409 Moscow, Russia\\
}

\email{*Ivan.Vartaniants@desy.de, Wilfried.Wurth@desy.de} 



\begin{abstract}
  The experimental characterization of the spatial and temporal coherence properties
  of the
  free-electron laser in Hamburg (FLASH) at a wavelength of 8.0 nm is presented. Double pinhole diffraction patterns of
  single femtosecond pulses focused to a size of about 10$\times$10 $\mu$m$^2$ were measured. A transverse
  coherence length of $6.2\pm0.9~\mu$m
  in the horizontal and $8.7\pm1.0~\mu$m in the vertical direction was determined from the most coherent pulses.
  Using a split and delay unit the coherence time of the pulses
  produced in the same operation conditions of FLASH was measured to be $1.75\pm0.01$ fs. From our experiment
  we estimated the degeneracy parameter of the FLASH beam to be on the order of $10^{10}$ to $10^{11}$, which exceeds the values of this parameter at any other source in the same energy range by many orders of magnitude.
\end{abstract}


\section{Introduction}
Free-electron lasers (FELs) based on the self-amplified spontaneous
emission (SASE) principle produce extremely brilliant, highly coherent
radiation in the extreme ultraviolet (XUV) \cite{A2007} and hard x-ray \cite{E2010} range. Utilizing the high photon flux,
the femtosecond pulse duration and
the high degree of coherence, techniques like coherent x-ray
diffraction imaging (CXDI) \cite{MCK1999,PWV2006,C2006,N2010}
x-ray holography \cite{ELS2004} and, recently,
nano-crystallography \cite{C2011} promise important new insights in
biology \cite{SEM2011,MYV2010}, condensed matter physics
\cite{V2007} and atomic physics \cite{Y2011}.
Some of these methods can be implemented only if
the radiation is sufficiently coherent, both spatially and temporally.
This means that the knowledge of the
coherence is mandatory for the success of these experiments.
Moreover, it was shown recently that
small deviations from perfect coherence can be taken into
account in the CXDI method if
the degree of coherence is known \cite{WWQ2009,AWQ2011}.
Both, temporal and transverse coherence effects also play a role in the
now accessible field of non-linear excitations of atoms and molecules \cite{JY2010}.

Free-electron laser in Hamburg (FLASH) started its operation in 2005 as the first FEL in the XUV wavelength range. It delivered radiation of unprecedented brightness at a wavelength down to {6.5 nm} (first reached in October 2007)
and generated pulses as short as 10-15 fs Full Width at Half Maximum (FWHM) \cite{A2007}. A variety of ground breaking
experiments \cite{FLASH}, including the first demonstration of single pulse femtosecond coherent imaging \cite{C2006} and studies of the photoelectric effect at ultra-high intensities \cite{SBF2007},
were performed at FLASH. In 2009 FLASH went through a major upgrade \cite{HFF2010},
where, along with another electron energy upgrade yielding photon wavelengths down to about 4 nm,
a 3rd harmonic accelerator module was installed, in order to improve the capabilities for
longitudinal electron bunch compression and in this way enhance the machine stability.
This led to a higher stability of the FEL operation at the expense of longer pulse durations, typically around 100 fs.
The aim of our experiment was to characterize the coherence properties of the FLASH beam after this upgrade.

Measurements of transverse coherence properties of FEL sources have been reported earlier
\cite{SVK2008,VMS2010,RSW2011}.
In these experiments a number of shots
were averaged and an average transverse coherence length was determined.
Contrary to a visible laser, where a resonator typically permits the growth of only a single
transverse ''TEM$_{00}$'' mode, in a SASE FEL a variety of modes can be amplified \cite{SE2010}.
The signal at the end of the undulator
depends on the shot noise in the electron bunch entering the undulator. As such, the radiation
properties, including the transverse coherence, may change from shot to shot.
Although no significant shot to shot variations of the coherence properties were
observed at the Linac Coherent Light Source (LCLS) \cite{VSM2011}, at FLASH, with its lower electron and
photon energies, we may expect these variations to be more important. In this paper we present
measurements of the transverse coherence properties of FLASH and investigate how they fluctuate from shot to shot.
To address this important question, we employed the single-shot methodology developed in \cite{VSM2011} and
conducted Young's double pinhole experiment \cite{G2000,MW1995} with single FEL pulses.

The other important statistical characteristic of the FEL radiation is its temporal coherence. Due to the same
electron bunch instabilities, which lead to partial spatial coherence, the FEL pulses
are partially coherent in the time domain. The photon pulse is not merely a single longitudinal
mode but rather consists of several spikes with a width of about the
coherence time. These longitudinal modes can be correlated and
interfere with each other, which affects the coherence time. This complicated phenomena was recently observed
experimentally \cite{SW2010}. To analyze the temporal coherence properties
at FLASH we used an autocorrelation setup \cite{SF2010}. The single FLASH pulse
was split into two parts on a wavefront dividing split mirror. These two parts were brought to an overlap on the detector with an adjustable time delay. The visibility of the interference fringes in the overlap region contains information on the degree of correlation of the time delayed pulses, and therefore on the temporal coherence properties of the radiation.

%
A number of measurements of the transverse \cite{SVK2008,VMS2010,RSW2011} and
temporal \cite{RSW2011,MR2009,SW2010} coherence of FLASH operating before the upgrade have been reported. Here
we combine spatial and temporal coherence measurements using Young's double pinholes
and a split and delay unit at the same operation conditions of FLASH after its upgrade in 2009.

\section{Theory}
The concept of optical coherence is associated with interference phenomena, where the Mutual Coherence
Function (MCF) \cite{G2000,MW1995}
\begin{equation}
  \Gamma_{12}(\tau) = \langle E^*(\mathbf r_1,t)E(\mathbf r_2,t+\tau)\rangle_T
  \label{eq:Gamma}
\end{equation}
plays the major role. It describes the
correlation between two complex values of the electric field $E^*(\mathbf r_1,t)$ and
$E(\mathbf r_2,t+\tau)$ at different points $\mathbf r_1$ and $\mathbf r_2$ in space, separated by the time interval
$\tau$. The brackets $\langle\cdots\rangle_T$ denote the time average over a time interval $T$.

To experimentally characterize the MCF, the spatial and temporal properties are
measured. The former can be accessed by Young's double pinhole
experiment \cite{G2000,MW1995},
whereas the latter may be studied using a split and delay line \cite{SW2010,MR2009}.
In these interference experiments
the complex field $E(\mathbf r,t)$ is divided
into two parts, $E_1(\mathbf r_1,t)$ and $E_2(\mathbf r_2,t)$, by the double pinhole or the split mirror. Here
$\mathbf r_1$ and $\mathbf r_2$ are the positions in the plane of the apertures or the split mirror.
The interference between the two propagated fields $E_1(\mathbf u,t)$ and $E_2(\mathbf u,t+\tau)$ is observed
in the overlap region
on the detector and the degree of coherence of the
incident radiation field can be determined from the contrast of the
interference fringes. The coordinate $\mathbf u$ lies in the observation plane and $\tau$ is the time delay, which
is introduced through the difference between the propagation path lengths
of the two beams.
For narrowband light with an average wavelength $\lambda$
the intensity distribution measured in the observation plane may be expressed as \cite{G2000}
\begin{equation}
  I(\mathbf u)=I_1(\mathbf u)+ I_2(\mathbf u)+
  2\sqrt{I_1(\mathbf u) I_2(\mathbf u)}
  \left|\gamma_{12}(\tau)\right|\cos\left[\delta(\mathbf u)+\alpha_{12}(\tau)\right],
  \label{eq:ds_general}
\end{equation}
where $I_{1,2}(\mathbf u)$ are the intensities of the individual fields and
\begin{equation}
  \gamma_{12}(\tau) = \frac{\Gamma_{12}(\tau)} {\sqrt{I_1 I_2}}
                    = \left|\gamma_{12}(\tau)\right| e^{i[\delta(\mathbf u)+\alpha_{12}(\tau)]}
  \label{eq:gamma}
\end{equation}
is the complex degree of coherence in the plane of the apertures or the split mirror. Here, according to
the definition (\ref{eq:Gamma}) $I_1\equiv\Gamma_{11}(0)$, $I_2\equiv\Gamma_{22}(0)$ are the intensities incident
on the double pinhole or on the split mirror.
In equation (\ref{eq:ds_general}) $\delta(\mathbf u)$ is the rapidly changing phase of $\gamma_{12}(\tau)$,
which gives rise to interference
fringes in the observation plane. The slowly varying phase of $\gamma_{12}(\tau)$ is denoted by $\alpha_{12}(\tau)$.

In a Young's double pinhole experiment
the fast varying phase term $\delta(\mathbf u)$ can be
expressed as \cite{G2000} $\delta(\mathbf u)=2\pi\mathbf u\cdot\mathbf d/(\lambda z)$.
Here $z$ is the double pinhole to detector distance,
and $\mathbf d=(d_x,d_y)$ is the pinhole separation in
the horizontal $d_x$ and vertical $d_y$ direction. The intensities $I_{1,2}(\mathbf u)$  in equation (\ref{eq:ds_general})
are given by $I_{1,2}(\mathbf u)=I_{1,2} I_D(\mathbf u)$, where
$I_D(\mathbf u)$ is the Airy diffraction pattern from a round pinhole of diameter $D$ \cite{G2000}.
The incident intensity is assumed to be constant across each pinhole. For sufficiently big pinhole to detector distances the Airy patterns
from different pinholes overlap on the detector \cite{G2000} and equation (\ref{eq:ds_general}) can be simplified to
\begin{equation}
  I(\mathbf u)=(I_1+I_2)I_D(\mathbf u)
  \left\{1+\left|\gamma_{12}^{\rm eff}(\tau)\right|\cos\left[\frac{2\pi}{\lambda z}\mathbf u\cdot \mathbf d+\alpha_{12}(\tau)\right]\right\},
  \label{eq:fit}
\end{equation}
where the effective complex degree of coherence is given by
\begin{equation}
  \gamma_{12}^{\rm eff}(\tau)=\frac{2\sqrt{I_1I_2}}{I_1+I_2}\gamma_{12}(\tau).
  \label{eq:gamma_eff}
\end{equation}
From the definition of $\gamma^{\rm eff}_{12}(\tau)$ in equation (\ref{eq:gamma_eff})
it is immediately seen that if the intensities incident on
individual pinholes differ in magnitude, the contrast
observed in the experiment is reduced as compared with the degree of coherence of the incident radiation. If the
time delay $\tau$ associated with the path length difference between the pinhole one and two is smaller than the coherence time $\tau_c$, then
$\gamma_{12}^{\rm eff}(\tau) \approx \gamma_{12}^{\rm eff}(0)$ and $\alpha_{12}(\tau) \approx \alpha_{12}(0)$ are
good approximations. Typically, the double pinhole experiment is conducted for different pinhole separations $\mathbf d$
and the transverse coherence length is
defined as a characteristic width of $|\gamma_{12}(0)|$. To
characterize the transverse coherence it is convenient to
introduce the normalized degree of coherence as \cite{SSY2008,VS2010}
\begin{equation}
  \zeta =
  \frac{\int |\Gamma_{12}(0)|^2\mbox d \mathbf r_1\mbox d \mathbf r_2 }
  {\left(\int\Gamma_{11}(0)\mbox d\mathbf r_1\right)^2 }.
  \label{eq:NDC}
\end{equation}
This quantity approaches unity for highly coherent and zero for incoherent radiation.


In the split and delay line the interfering fields are generated at the split mirror and are propagated
through different arms of the instrument. Both beams
are brought to full overlap in the observation plane with
approximately equal intensities $I_1(\mathbf u)\approx I_2(\mathbf u)$.
The interference fringes occur due to the recombination angle  $\theta$,
which also is the angle between the wavefronts of both beams.
The fast varying phase term in equation (\ref{eq:ds_general})
in this case is given by \cite{RSW2011}
$\delta(\mathbf u)=2\pi \mathbf u\cdot\mathbf n_\perp\tan(\theta) /\lambda$, where $\mathbf n_\perp$
is the unit vector along the direction
perpendicular to the interference fringes.
The analysis of the visibility of fringes can be conveniently performed using the Fourier transform.
Assuming that the modulus of the complex coherence function $|\gamma_{12}(\tau)|$ and
$\alpha_{12}(\tau)$ are constant within
the interference pattern, the Fourier transform of equation (\ref{eq:ds_general}) yields \cite{BPG2002}
\begin{equation}
  \tilde I(\mathbf f) = 2\tilde I_1(\mathbf f)\otimes
  \left\{\delta(\mathbf f) + \frac12 |\gamma_{12}(\tau)|
  \left[\delta(\mathbf f+\mathbf f_s)e^{-i\alpha_{12}(\tau)}
  + \delta(\mathbf f-\mathbf f_s)e^{i\alpha_{12}(\tau)} \right]\right\},
  \label{eq:sdl_ft}
\end{equation}
where $\tilde I(\mathbf f)$ and $\tilde I_1(\mathbf f)$ denote the Fourier transforms of $I(\mathbf u)$ and
$I_1(\mathbf u)$, $\delta$ is the Dirac delta function, $\otimes$ is the convolution operator,
$\mathbf f$ is the frequency coordinate, and the fringe frequency is given by
$\mathbf f_s = \mathbf n_\perp\lambda /\tan(\theta)$. Equation (\ref{eq:sdl_ft})
contains one central term and two 'sideband' terms.  The central term is the sum of
the autocorrelation functions of the two beams. The 'sideband' terms
contain the information on the degree of cross correlation
between the interfering fields.
According to equation (\ref{eq:sdl_ft}) the modulus of the complex coherence function can be determined as
\begin{equation}
  |\gamma_{12}(\tau)|=2\frac{|\tilde I(\mathbf f_s)|}{|\tilde I(0)|}.
  \label{eq:sdl_ft1}
\end{equation}

For a fixed overlap the visibility of the resulting fringes is analyzed as a function of the time delay $\tau_c$
caused by the two different propagation distances in the split and delay unit. From these measurements one can
deduce the coherence time of the radiation, $\tau_c$, which we defined according to
\cite{G2000} as
\begin{equation}\label{eq:length}
  \tau_c=\int |g(\tau)|^2 \mbox d\tau,
\end{equation}
where
\begin{equation}
  \label{eq:normalizedgamma}
  g(\tau)=\frac{\gamma_{12}(\tau)}{\gamma_{12}(0)}.
\end{equation}

In FEL theory the coherence time can be approximated by \cite{SE2010,MB2010}
\begin{equation}
\label{eq:coherencetime}
\tau_c^{\rm th}\approx \frac{\lambda}{4\pi\rho c},
\end{equation}
where $\lambda$ is the resonance frequency,
$\rho$ is the FEL  parameter and $c$ is the speed of light.
The FEL parameter in equation (\ref{eq:coherencetime}) can be expressed as
$\rho=[IA_{JJ}^2K^2\lambda_{\rm u}^2/(32\pi^2 I_A \gamma^3\sigma_{\bot}^2)]^{1/3}$,
where $I$ is the electron beam current, $K$ is the undulator parameter,
$\lambda_{\rm u}$ is the undulator period, $I_A=17~kA$ is the Alfv\'en current,
$\gamma$ is the relativistic (Lorentz) factor, and $\sigma_{\bot}$
is the transverse Root Mean Square (rms) size of the electron bunch. The coupling factor
$A_{JJ}$ for a planar undulator is $A_{JJ} =|J_0(Q)-J_1(Q)|$,
where $J_0$ and $J_1$ are Bessel functions of the first kind and
$Q=K^2/[2(1+K^2)]$.

\section{Experiment}
\begin{figure}[t]
  \centering\includegraphics[width=0.6\textwidth]{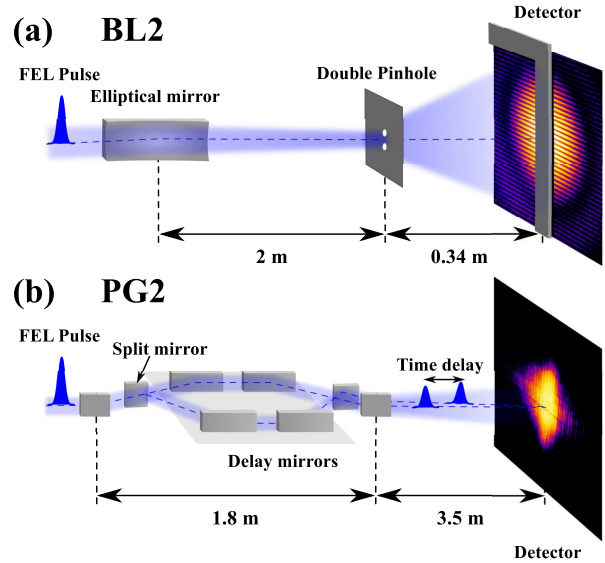}
\caption{A sketch of the experiment.
a) Young's double pinhole experiment to characterize the spatial coherence of the incident beam at the BL2 beamline.
b) The split and delay unit at the PG2 beamline splits the incoming beam into two beams. Both beams are
recombined in space on the detector. Due to the geometry of the setup both beams overlap with a small angle.
\label{fig:sketch}
}
\end{figure}
The transverse and temporal coherence of FLASH were measured at a lasing wavelength of
8.0 nm.
The FEL generated single pulses at a repetition rate of 10 Hz with an average energy of 180 $\mu$J per pulse.
FLASH was operated with a bunch charge of
0.8 nC and an electron energy of around 891 MeV. Six modules of the undulator with a total length of 30 m
were used. The FEL was expected
to lase in the saturation regime at these operation conditions.
The pulse duration was estimated to be about 100 fs (see below).
The outline of both experiments for the measurements of the transverse and longitudinal coherence
is shown in Figure \ref{fig:sketch}.

\subsection*{(a) Transverse coherence measurements}
The spatial coherence measurements were carried out at the BL2 beamline
at FLASH (see Figure \ref{fig:sketch}(a)).
The beam delivery system consisted of two flat distribution mirrors and an
elliptical mirror, which focused the beam to a size of about $10 \times 10$ $\mu m^2$ (FWHM) 70 m
downstream from the undulator exit.
The acceptance of the mirrors was sufficiently large in both
directions, therefore we assume that the beam was not cut by the mirrors.
The double pinhole apertures were positioned in the focus inside a dedicated vacuum chamber (HORST \cite{S2009}).
Double pinholes with varying separations between the pinholes have been manufactured to measure the degree
of coherence at different relative distances. The pinhole separation varied between 4 $\mu$m and  11 $\mu$m in the horizontal direction and between 2 $\mu$m and 15 $\mu$m in the vertical direction. With increasing pinhole separation less intense parts of the beam were probed. To record a similar signal from all apertures we varied the pinhole diameter between 340 nm for the smallest pinhole separation and 500 nm for the largest pinhole separation.

Due to the extremely high power densities in the focus, each aperture set was destroyed during the interaction with a single
FEL pulse. About ten identical apertures were manufactured for each pinhole separation to improve statistics
of the measurements. All apertures were fabricated by electroplating a 1.3 $\mu$m thick gold layer on top
of a 100 nm silicon nitride substrate. The substrate
was supported by 50$\times$50 $\mu m^2$ windows etched in a 200 $\mu$m thick silicon wafer
and was eventually removed by hydrofluoric acid to increase the transmission through the pinholes.
Individual pairs of apertures were separated by 768 $\mu$m from each other in both directions.

The double pinhole diffraction patterns were recorded with an in-vacuum Charge Coupled Device (CCD) (LOT/Andor DODX436-BN)
with $2048\times 2048$ pixels, each $(13.5~\mu$m$)^2$ in size.
A 3 mm linear beamstop manufactured out of B$_4$C was oriented perpendicular to the interference fringes
and protected the CCD from the direct FEL beam (see Figure \ref{fig:sketch}(a)). A 200 nm thick Pd foil supported by 100 nm parylene-N was mounted $29$ mm upstream from the camera
to absorb the visible light generated during the damage process of the apertures.
A sample to detector distance of 0.34 m provided a sufficient sampling of the fringes (13 pixels per fringe for a 15 $\mu$m pinhole separation),
which is necessary for the evaluation of the double pinhole interference patterns.

\subsection*{(b) Temporal coherence measurements}
\begin{figure}[t]
  \centering\includegraphics[width=0.6\textwidth]{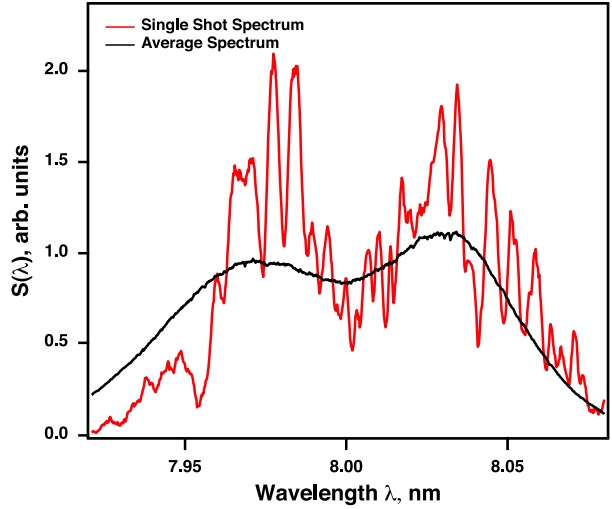}
\caption{A typical single pulse (red line) and average (black line) spectrum of the FLASH radiation.
}
\label{fig:spectrum}
\end{figure}
The temporal coherence measurements were carried out directly after the transverse coherence measurements  at the plane grating monochromator beamline PG2 \cite{WM2007}. There was no further tuning of the machine involved.
A typical single pulse and average spectra are shown in Figure \ref{fig:spectrum}.
From the average spectrum we estimate a FWHM bandwidth
of about 1.4\% for the FEL radiation. Analyzing individual single shot spectra we estimate an average pulse duration
to be on the order of $100$ fs. This number is retrieved by counting the individual spikes in the single
shot spectra and multiplying the average number of spikes per pulse with the measured coherence time (see below).

The plane grating monochromator was used in zero order, and the longitudinal coherence was measured by using the permanently installed split and delay unit \cite{SF2010} (see Figure \ref{fig:sketch}(b)) of the PG2 beamline. This device is able to
geometrically split each pulse delivered by FLASH into two pulses and delay them up to 5.1 ps in time with respect
to each other with less than 100 as accuracy. Afterwards, the two pulses can be overlapped in space on a Ce:YAG screen 3.5 m downstream in the beamline creating interference fringes when the pulses overlap in time.
The images of the overlapped pulses were recorded at the machine pulse repetition rate of 10 Hz
using a Basler scA1300-32fm FireWire CCD camera outside of the vacuum chamber. The camera has in
total $1296\times 966$ pixels with a pixel
size of $(3.75~\mu$m$)^2$. As optics we used a Sill Optics S5LPJ0635 lens with a magnification
of $1.102$ yielding sufficient resolution to resolve the interference fringes. The complete setup
was mounted on an x-y positioning stage allowing for a fast alignment of the camera to the beam position.
The measurements were not done in the focus of the beam but rather between the two intermediate
foci for the horizontal and the vertical direction.
\section{Results}
\subsection*{(a) Transverse coherence measurements}
\begin{figure}[t!]
\centering
\includegraphics[width=0.8\textwidth]{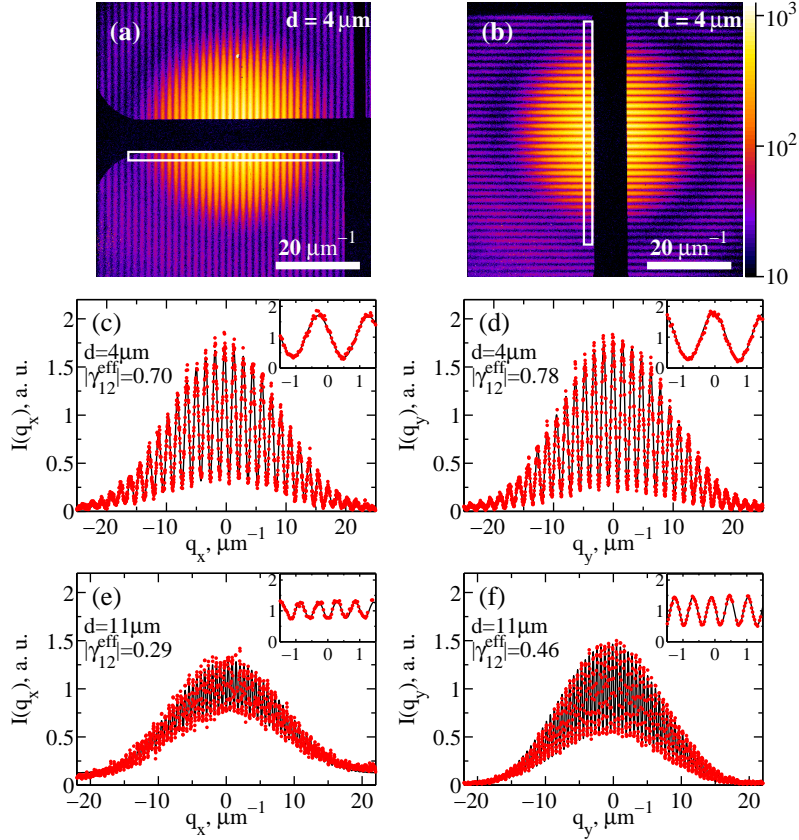}
\caption{The double pinhole diffraction patterns measured with pairs
of pinholes separated by $4~\mu$m oriented horizontally (a)
and vertically (b) shown in logarithmic scale.
The analyzed region is indicated by the white rectangle.
(c,d) Line scans from the analyzed region close to the center of the diffraction patterns (a,b) (red dots)
and the theoretical fit (black line).
(e,f) The same as in (c,d) for $11~\mu$m pinhole separation. The insets in (c-f) show enlarged regions of the
respective curves.
}
\label{fig:DP1}
\end{figure}
Typical recorded and dark field corrected single shot diffraction patterns are shown in Figure \ref{fig:DP1}(a,b)
as a function of the momentum transfer $\mathbf q$ for a pinhole separation of 4 $\mu$m.
In Figure \ref{fig:DP1}(a) the double pinhole was
oriented horizontally and the vertical fringes originate from the interference between
the field scattered at different pinholes. In Figure \ref{fig:DP1}(b) a similar diffraction pattern measured
with a vertically oriented double pinhole is presented. The first minimum of the Airy distribution is visible at
$|\mathbf q|\approx 25~\mu$m$^{-1}$ in both figures.
The line scans through the measured diffraction patterns
(Figure \ref{fig:DP1}(c,d)) show a high contrast level for
the small pinhole separation for both, the horizontal and vertical directions.
The contrast decreases for larger separations (see Figure \ref{fig:DP1}(e,f)), which indicates
a smaller magnitude of the complex degree of coherence at these length scales.

On most of the diffraction patterns additional noise was observed. It consisted of a constant background and
a few hot pixels randomly distributed over the whole diffraction pattern ('salt and pepper noise').
We attribute the appearance of this noise to the light generated during the
damage process of the pinholes. Since the Pd foil was not attached to the detector but was positioned a few centimeters upstream, light could leak between the foil and detector and can be a source of this noise.

To determine the modulus of the effective complex degree of coherence $|\gamma_{12}^{\rm eff}|$
for each measured single shot interference pattern, equation
(\ref{eq:fit}) was fit to the data \cite{footnote1}.
We added a constant $A$ to equation (\ref{eq:fit}) to accommodate for the presence of the constant background
noise. The hot pixels were removed from the diffraction patterns and were not considered in the analysis
procedure.
In particular, the two-dimensional area marked with a white rectangle in Figure \ref{fig:DP1}(a,b) was analyzed.
Eight fit parameters including
$\left|\gamma_{12}^{\rm eff}\right|$, $(I_1+I_2)$, $D$, $d$, $\alpha_{12}$, $A$ and
the position of the optical axis in the horizontal
and vertical directions were found. The quality of fit was characterized by an $R$-factor
$R=\sum_i(I^{\rm th}_i-I^{\rm exp}_i)^2/\sum_i(I^{\rm exp}_i)^2$, where
$I^{\rm exp}$ is the background corrected measured data, $I^{\rm th}$ is the fit and summation is made over all points in the fitted area.
All fits with $R>0.01$ were excluded from the further analysis (less than 50\%
from the total number of the diffraction patterns in each direction).
For each shot a confidence interval of $|\gamma_{12}^{\rm eff}|$ was determined as a value for which $R$ was twice as large as the minimum value,
while all other fit parameters were fixed. Typical fit results are shown in Figure \ref{fig:DP1}(c-f).

\begin{figure}[t!]
\centering
\includegraphics[width=0.8\textwidth]{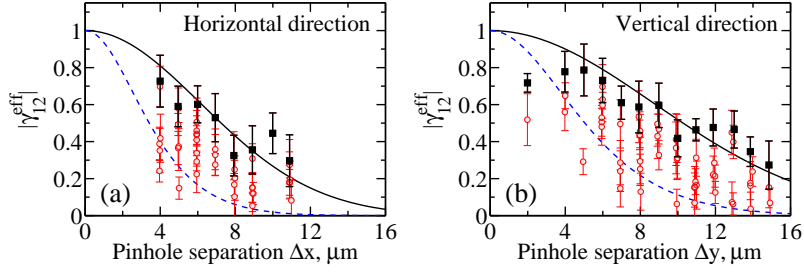}
\caption{The modulus of the effective complex degree of coherence in horizontal (a) and vertical (b) direction. The red
circles and black squares show the measured values. Each point corresponds to a single FEL shot. Gaussian fits (black lines) through the highest values (black squares) gave for the coherence length the following result $l_c^H=6.2\pm0.9~\mu$m in the horizontal and $l_c^V=8.7\pm1.0~\mu$m in the vertical direction for a beam size of $(10\pm2)\times(10\pm2)~\mu$m$^2$ FWHM. Blue dashed lines show the reduction of contrast due to beam positioning instabilities (see text).}
\label{fig:DP2}
\end{figure}

As a result of the data analysis, the modulus of the effective complex degree of coherence
$|\gamma_{12}^{\rm eff}|$ as a function of the pinhole separation is shown
in Figure \ref{fig:DP2}(a,b) for the horizontal and vertical directions. We approximated the highest
values of $|\gamma^{\rm eff}_{12}|$ for each pinhole separation (shown by black squares)
with the Gaussian function $\exp[-d^2/(2l_c^2)]$ shown by black line in Figure \ref{fig:DP2}(a,b). This yields an upper bound estimate of the transverse coherence length in each direction. In this way we determined the transverse coherence length (rms)
to be $l_c^H=6.2\pm0.9~\mu$m in the horizontal and $l_c^V=8.7\pm1.0~\mu$m in the vertical direction. During this fitting
procedure we fixed the value of $|\gamma_{11}^{\rm eff}|$ at zero pinhole separation to one according
to its definition (\ref{eq:Gamma},\ref{eq:gamma},\ref{eq:gamma_eff}) \cite{footnote4}.

According to equation (\ref{eq:gamma}) the information about the intensity profile of the beam is required to
characterize the MCF $\Gamma_{12}(\tau)$ completely.
To measure the beam profile in the plane of the apertures we analyzed
PMMA imprints produced by single FEL pulses with a varying degree of attenuation of the beam \cite{CKJ2010}.
Three sets of PMMA imprints with one order of magnitude difference in attenuation of the incoming beam were analyzed.
Using the Gaussian beam approximation a beam size of $(10\pm2)\times(10\pm2)~\mu$m$^2$ FWHM was determined.
In the horizontal direction additional features on the sides of the beam were observed.
For the strongly attenuated beam,
however, round craters, $15~\mu$m in diameter, indicate that the central part of the beam is round.

As follows from our analysis (see Figure \ref{fig:DP2}) the values of
$|\gamma_{12}^{\rm eff}|$ vary significantly from shot to shot for the same pinhole separation. We attribute
this variation mainly to the beam position instabilities in the plane of the sample.
If the center of the beam does not hit the
center of the double pinhole the intensities incident on individual pinholes will necessarily be different. This difference
yields a reduced value of $|\gamma_{12}^{\rm eff}|$ as compared with $|\gamma_{12}|$ (see equation (\ref{eq:gamma_eff})).
We estimated the deviation of the beam center relative to the sample by analysing the PMMA imprints measurements.
The positions of thirty two craters in the PMMA were found and compared
with the nominal positions expected from the stage movement.
A maximum offset between the position of the apertures and the incident
beam was determined to be
$\pm12~\mu$m in the horizontal and $\pm8~\mu$m in the vertical direction
\cite{footnote5}.
Using these values as the offset of a
Gaussian beam with a size of $10\times10~\mu m^2$, we can calculate the difference of the intensity
incident on pinhole one and two. The error imposed by this uncertainty in position compared to the Gaussian fit through
the highest values of $|\gamma_{12}^{\rm eff}|$ (black solid line in Figure \ref{fig:DP2}(a,b)) is shown by the blue
dashed line in Figure \ref{fig:DP2}(a,b). Most of the measured values lie between the black and the blue
line, which indicates that the positional uncertainty is the dominant cause for the apparent variations in
$|\gamma_{12}^{\rm eff}|$. However, as this error is
quite significant, we cannot definitively exclude
shot to shot variations in the degree of coherence $|\gamma_{12}|$.

Combining the measured beam size
with the highest values of the complex coherence function we
determined the degree of transverse coherence $\zeta$ (see equation (\ref{eq:NDC})) of the radiation at FLASH.
Utilizing the Gaussian Schell-model \cite{MW1995}
we estimated a value of $\zeta_x=0.59\pm0.10$ in the horizontal
and $\zeta_y=0.72\pm0.08$ in the vertical directions. That gives for the
total degree of coherence obtained in our experiment a value of $\zeta=\zeta_x\zeta_y=0.42\pm0.09$.
The decomposition of the radiation field into
a sum of coherent modes \cite{MW1995, VS2010} has shown that 2 modes in each direction
are sufficient to describe the coherence properties in the measured area of the beam.
This means (see for derivation \cite{VSM2011}) that about $62\pm11$\% of the total radiation
power is concentrated in the dominant transverse mode.

The coherence measurements presented here indicate a significantly higher degree
of transverse coherence of the FLASH beam than previously reported values \cite{SVK2008}. We attribute this
to the implementation of the 3rd harmonic cavity to the FLASH accelerator complex \cite{HFF2010}. A comparison
with values reported for the LCLS \cite{VSM2011} shows, that both machines, though operating at significantly different
wavelengths and different pulse energies, provide similar values of the degree of coherence
(in the measurements at LCLS \cite{VSM2011} the wavelength was 1.6 nm, the average energy per pulse 1 mJ and the
degree of coherence about 75\%).


\subsection*{(b) Temporal coherence measurements}
The visibility of the interference pattern was measured as a function of the time delay between the two pulses. From these
measurements one can deduce the mean electric field autocorrelation function of the FEL radiation.
The analysis of the data was done by performing a two-dimensional Fourier transform of the recorded
fringe patterns (see equation (\ref{eq:sdl_ft}) and reference \cite{SW2010} for details). In comparison with the experiment
reported in \cite{SW2010} the detector unit has been significantly
improved by isolation of the detection system from the background light
(light from the experimental hall) and magnification of the FEL beam size on the detector.
Since FLASH was operating in single bunch mode, each image contains the interference pattern
of a single pulse on the Ce:YAG crystal.
Therefore, the presented data was not affected by blurring of the interference patterns
and thus reduction of the contrast.
This blurring can appear for
long delays when microbunches in a bunch train have a slightly different wavelength which leads to a change of the
fringe spacing \cite{SW2010}.

\begin{figure}[t]
\centering
\includegraphics[width=0.8\textwidth]{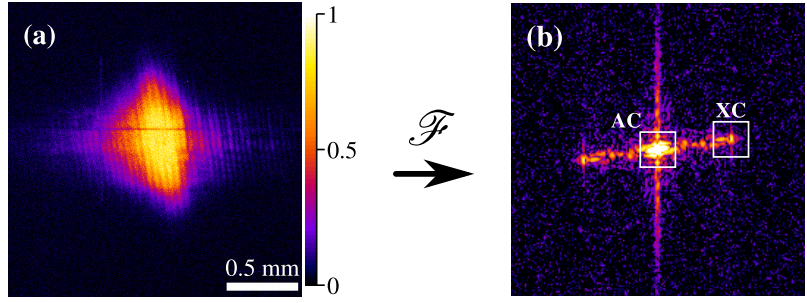}
\caption{(a) Typical background subtracted single pulse interference fringes
in the overlap region of the two beams in space and time
recorded on the detector. (b) The magnitude of the Fourier transform of the interferogram (a).
The AC marked region is equivalent to the overall intensity in the image while
the XC marked part is the amplitude of the fringe pattern.}
\label{fig:rawdata}
\end{figure}

A typical single shot interferogram for the time delay $\tau=0$ is shown in Figure~\ref{fig:rawdata}(a).
From the recorded interference patterns we subtracted the background, which was calculated
for each image by averaging a region in the corner of the image. After this background subtraction
the occasionally occurring negative values were set to zero.
A two dimensional Fourier transform (see Figure \ref{fig:rawdata}(b))
of the processed image was calculated and also background corrected.
The modulus of the complex coherence function $|\gamma_{12}(\tau)|$
was found for each single shot according to equation (\ref{eq:sdl_ft1}). An average of the Fourier transform of the data in the regions marked by AC (autocorrelation)
and XC (cross correlation) (see Figure \ref{fig:rawdata}(b)) were used in this analysis.
A number of single shot measurements for each
time delay $\tau$ was averaged to improve the statistics and to determine the average complex coherence function
as a function of the time delay \cite{footnote3}.
%

\begin{figure}[t]
\centering
\includegraphics[width=0.8\textwidth]{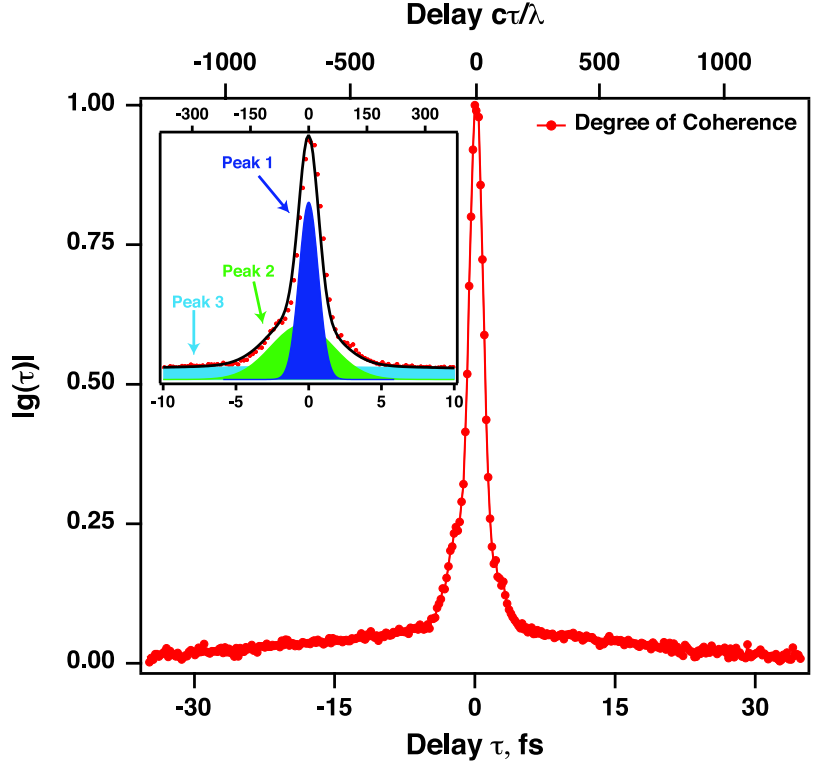}
\caption{The modulus of the complex coherence function $g(\tau)$ as a function
of the time delay. The top and bottom axes show the delay in wavecycles and femtoseconds, respectively.
In the inset an enlarged region of $\pm10$ fs, which corresponds to about $\pm375$ wavecycles, is shown.
Three timescales can be observed. One short coherence peak, an intermediate asymmetry component and a
long tail. The red dots are the measured data and the black curve is the sum of the three Gaussian functions.}
\label{fig:zeroordercoherence}
\end{figure}

\begin{table}[b]
\caption{Full width at half maximum of the three time scales contributing to the temporal coherence as determined by fitting three Gaussian functions to the data.}
\begin{center}
\label{tab:comparison}
\begin{tabular}{| c | c | c |}
\hline
	 Peak \# & FWHM (fs)& FWHM $(c\tau/\lambda)$\\
	\hline
	\hline
	Peak 1& $1.54\pm0.01$ & $57.8\pm0.5$\\
	\hline
	Peak 2& $4.93\pm0.06$ & $185\pm2$\\
	\hline
	Peak 3& $42\pm1$ & $1575\pm37$\\
	\hline
\end{tabular}
\end{center}
\end{table}

In order to determine the coherence time the modulus of the complex coherence function
was normalized according to equation (\ref{eq:normalizedgamma}). Figure \ref{fig:zeroordercoherence} shows
the normalized value $|g(\tau)|$ as a function of the time delay.
The measured coherence time determined by equation (\ref{eq:length}) has a value of
$\tau_c^{\rm exp}=1.75\pm0.01$ fs, which corresponds to about $65.5\pm0.5$
wavecycles ($c\tau/\lambda$).
For the FLASH parameters used in our experiment \cite{parameters} we estimated the FEL parameter to be $\rho=1.9\times10^{-3}$.
That gave theoretical estimate of the coherence time using equation (\ref{eq:coherencetime}) $\tau_c^{\rm th}=1.1\pm0.3$ fs, that concords well with the measured coherence time.

Close inspection of Figure \ref{fig:zeroordercoherence} shows, that the modulus of the complex coherence function
contains several time scales, which has been also reported in \cite{SW2010}.
To determine these time scales the data were fitted with three Gaussian functions.
The result of this fit is presented in Table 1. We interprete these three contributions in the following way.
The shortest peak 1 with a width of $1.54\pm0.01$ fs ($57.8\pm0.8$ wavecycles) is the contribution of the coherence time of the single FEL spikes within one pulse. The longest peak 3 with a width of $42\pm1$ fs ($1575\pm37$ wavecycles) describes the decay of the degree of correlation between individual spikes in the FEL pulse.
In our experiment, the normalized temporal coherence reaches a value close to zero for delays of about $1300$ wavecycles, whereas the measurement of the temporal coherence reported in \cite{SW2010} at 9.6 nm has reached zero already at about $300$ wavecycles. In this experiment, the electron bunch charge was set to 0.76 nC, which is close to the 0.8 nC, used during our measurements. However, before the upgrade, FLASH was operated with a non symmetrical electron bunch shape
yielding a leading high current peak and a long tail \cite{HFF2010}.
After the implementation of the 3rd harmonic cavity the pulses tend to be longer. Short pulses were not accessible during our measurements immediately following the restart of the FLASH after the upgrade.

The asymmetry, which can be seen in
Figure~\ref{fig:zeroordercoherence} for short delays ($\lessapprox 5$ fs) can be attributed to an asymmetry in the
two photon beams and is governed by peak 2 with a width of $4.93\pm0.06$ fs ($185\pm2$ wavecycles).
It may be a result of a non-constant non-linear chirp in the transverse direction of the photon beam, as well as a tilt of the wave front or longitudinal pulse double structures. Since we split the incoming beam by means of wavefront division, this can be back-translated to a non-linear chirp along the transverse direction of the photon beam. These asymmetries have also been measured in several other experiments using different techniques \cite{SW2010,MR2009,JY2010}.

\section{Discussion}
\begin{figure}[t]
  \centering\includegraphics[width=0.7\textwidth]{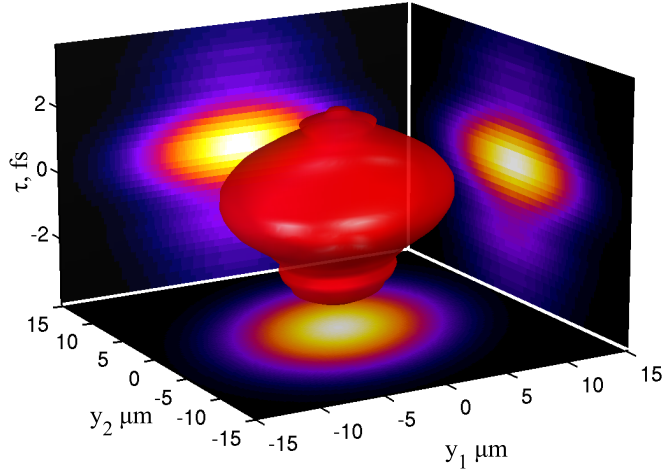}
\caption{A representation of the MCF $|\Gamma^V(y_1,y_2;\tau)|$ determined from the
measurements of the transverse (Gaussian fit to the maximum values) and temporal coherence properties.
The MCF was normalized according to
$|\Gamma^V(0,0;0)|=1$ and the isosurface with $|\Gamma^V(y_1,y_2;\tau)|=0.15$ (red),
is shown. The projections of $|\Gamma^V(y_1,y_2;\tau)|$ in $\tau$, $y_1$, $y_2$ directions are also presented.}
\label{fig:MCF}
\end{figure}

The statistical properties of the radiation at FLASH are described by the full mutual coherence function
$\Gamma(\mathbf r_1,\mathbf r_2;\tau)$.
We have characterized the MCF as a function of the space coordinates and as a function of the
time delay.
Combining the results from these measurements we determined the
magnitude of the complete MCF of the radiation at FLASH,
assuming the radiation is cross spectrally pure \cite{G2000}.
According to its definition (\ref{eq:Gamma}) the MCF is a function of two coordinates in space and one coordinate in
time. For visualization purposes we show a 3D representation of the MCF $|\Gamma^V(y_1,y_2;\tau)|$
in the vertical direction in Figure \ref{fig:MCF}.
A similar result is obtained for the MCF $|\Gamma^H(x_1,x_2;\tau)|$ in the horizontal direction.

Taking into account the pulse duration we can estimate the
degeneracy parameter $\delta$ \cite{MW1995,SE2010} of the FLASH beam.
It describes the number of photons
found in the same quantum state or a single mode of radiation.
The average total number of photons per single pulse was about
$7\times 10^{12}$.
From our transverse coherence measurements we
have approximated that about 65\% of the total power is concentrated in the dominant transverse mode. Using the
determined coherence time of about 2 fs and the average pulse duration of 100 fs we estimate
that about 1\% of the
total power is concentrated in a single longitudinal and transverse mode. This yields an estimate
of the degeneracy parameter $\delta \sim 10^{10}$ to $10^{11}$.
This number is significantly higher than at any other source at these photon energies.
For instance at synchrotron sources the degeneracy parameter is typically $\delta\le1$.
The measured degeneracy parameter of FLASH concords well with the theoretical predictions \cite{SE2010}
based on detailed SASE simulations. It is also similar to the degeneracy parameter of optical
light lasers \cite{MW1995}. This high value of the degeneracy could lead to new applications of FEL sources in the field of
quantum and non-linear optics in the XUV and x-ray regime.

\section{Conclusions}
We have experimentally characterized the transverse and longitudinal coherence properties of the
XUV free-electron laser FLASH. Young's double pinhole
experiment was conducted to find the transverse coherence length of the focused FEL beam. The
transverse coherence length of the focused FLASH beam was determined to be $6.2\pm0.9~\mu$m in the horizontal and
$8.7\pm1.0~\mu$m in the vertical direction.
Additionally, the intensity beam profile was measured to be $(10\pm2)\times(10\pm2)$ $\mu$m$^2$.
From our measurements we conclude that the focused FLASH beam is highly coherent with a total degree of transverse
coherence of about 40\%. A mode decomposition has shown that
about 60\% of the total power is concentrated in the fundamental mode. These high values indicate that almost
the full transverse photon flux is coherent and can be used for coherence-based applications.

The temporal coherence was measured to be $1.75\pm0.01$ fs, which is in good agreement with the expected theoretical
value of $1.1\pm0.3$ fs. While the main coherence peak fits well with the previous measurements, the broad component is about a factor of four larger. We attribute this effect to the longer pulses after the implementation of the 3rd harmonic cavity.

We have also estimated the degeneracy parameter of FLASH radiation to be in the range of
$10^{10}$ to $10^{11}$. This number is significantly larger than at any other existing sources operating at
this photon energy range and is comparable with the degeneracy parameter of conventional optical lasers.

\section{Acknowledgements}
We acknowledge the help of K. Hagemann in post processing the double pinhole samples for Young's experiments and
B. Faatz for his assistance in the theoretical estimates of the FEL properties. Part of this work was supported by BMBF Proposal 05K10CHG "Coherent Diffraction Imaging and Scattering of Ultrashort Coherent Pulses with Matter" in the framework of the German-Russian collaboration "Development and Use of Accelerator-Based Photon Sources".
We acknowledge the BMBF funding through the Virtual Institute VH-VI-403 from the Helmholtz society.
The KIT/Heidelberg group and the University of Hamburg acknowledges funding from BMBF (05K10VH4 and 05K10GU3, respectively within FSP 301 FLASH).

\begin{thebibliography}{99}
  \bibitem{A2007} W. Ackermann, G. Asova, V. Ayvazyan, A. Azima, N. Baboi, J. B\"ahr, V. Balandin, B. Beutner, A. Brandt, A. Bolzmann, R. Brinkmann, O. I. Brovko, M. Castellano, P. Castro, L. Catani, E. Chiadroni, S. Choroba, A. Cianchi, J. T. Costello, D. Cubaynes, J. Dardis, W. Decking, H. Delsim-Hashemi, A. Delserieys, G. Di Pirro, M. Dohlus, S. Düsterer, A. Eckhardt, H. T. Edwards, B. Faatz, J. Feldhaus, K. Flöttmann, J. Frisch, L. Fr\"ohlich, T. Garvey, U. Gensch, C. Gerth, M. G\"orler, N. Golubeva, H. J. Grabosch, M. Grecki, O. Grimm, K. Hacker, U. Hahn, J. H. Han, K. Honkavaara, T. Hott, M. H\"uning, Y. Ivanisenko, E. Jaeschke, W. Jalmuzna, T. Jezynski, R. Kammering, V. Katalev, K. Kavanagh, E. T. Kennedy, S. Khodyachykh, K. Klose, V. Kocharyan, M. K\"orfer, M. Kollewe, W. Koprek, S. Korepanov, D. Kostin, M. Krassilnikov, G. Kube, M. Kuhlmann, C. L. S. Lewis, L. Lilje, T. Limberg, D. Lipka, F. L\"ohl, H. Luna, M. Luong, M. Martins, M. Meyer, P. Michelato, V. Miltchev, W. D. M\"oller, L. Monaco, W. F. O. M\"uller, O. Napieralski, O. Napoly, P. Nicolosi, D. N\"olle, T. Nu\~nez, A. Oppelt, C. Pagani, R. Paparella, N. Pchalek, J. Pedregosa-Gutierrez, B. Petersen, B. Petrosyan, G. Petrosyan, L. Petrosyan, J. Pfl\"uger, E. Pl\"onjes, L. Poletto, K. Pozniak, E. Prat, D. Proch, P. Pucyk, P. Radcliffe, H. Redlin, K. Rehlich, M. Richter, M. Roehrs, J. Roensch, R. Romaniuk, M. Ross, J. Rossbach, V. Rybnikov, M. Sachwitz, E. L. Saldin, W. Sandner, H. Schlarb, B. Schmidt, M. Schmitz, P. Schm\"user, J. R. Schneider, E. A. Schneidmiller, S. Schnepp, S. Schreiber, M. Seidel, D. Sertore, A. V. Shabunov, C. Simon, S. Simrock, E. Sombrowski, A. A. Sorokin, P. Spanknebel, R. Spesyvtsev, L. Staykov, B. Steffen, F. Stephan, F. Stulle, H. Thom, K. Tiedtke, M. Tischer, S. Toleikis, R. Treusch, D. Trines, I. Tsakov, E. Vogel, T. Weiland, H. Weise, M. Wellh\"ofer, M. Wendt, I. Will, A. Winter, K. Wittenburg, W. Wurth, P. Yeates, M. V. Yurkov, I. Zagorodnov, and K. Zapfe, ''Operation of a free-electron laser from the extreme ultraviolet to the water window,'' Nat. Photonics \textbf{1}, 336-342 (2007).
  \bibitem{E2010} P. Emma, R. Akre, J. Arthur, R. Bionta, C. Bostedt, J. Bozek, A. Brachmann, P. Bucksbaum, R. Coffee, F. J. Decker, Y. Ding, D. Dowell, S. Edstrom, A. Fisher, J. Frisch, S. Gilevich, J. Hastings, G. Hays, P. Hering, Z. Huang, R. Iverson, H. Loos, M. Messerschmidt, A. Miahnahri, S. Moeller, H. D. Nuhn, G. Pile, D. Ratner, J. Rzepiela, D. Schultz, T. Smith, P. Stefan, H. Tompkins, J. Turner, J. Welch, W. White, J. Wu, G. Yocky, and J. Galayda, ''First lasing and operation of an angstrom-wavelength free-electron laser,'' Nat. Photonics \textbf{4}, 641-647 (2010).
  \bibitem{MCK1999} J. Miao, P. Charalambous, J. Kirz, and D. Sayre,
                    ''Extending the methodology of X-ray crystallography to allow imaging of micrometre-sized non-crystalline specimens,''
                    Nature \textbf{400}, 342 (1999).
  \bibitem{PWV2006} M. A. Pfeifer, G. J. Williams, I. A. Vartanyants, R. Harder and I. K. Robinson,
                    ''Three-dimensional mapping of a deformation field inside a nanocrystal,''
                    Nature \textbf{442}, 63 (2006).
                  \bibitem{C2006} H. N. Chapman, A. Barty, M. J. Bogan, S. Boutet, M. Frank, S. P. Hau-Riege, S. Marchesini, B. W. Woods, S. Bajt, W. H. Benner, R. A. London, E. Pl\"onjes, M. Kuhlmann, R. Treusch, S. D\"usterer, T. Tschentscher, J. R. Schneider, E. Spiller, T. M\"oller, C. Bostedt, M. Hoener, D. A. Shapiro, K. O. Hodgson, D. Van der Spoel, F. Burmeister, M. Bergh, C. Caleman, G. Huldt, M. M. Seibert, F. R. N. C. Maia, R. W. Lee, A. Sz\"oke, N. Timneanu, and J. Hajdu, ''Femtosecond diffractive imaging with a soft-X-ray free-electron laser,'' Nat. Phys. \textbf{2}, 839-843 (2006).
  \bibitem {N2010} K. A. Nugent,
                    ''Coherent Methods in the X-ray sciences'',
                    Advances in Physics \textbf{59}, 1-99 (2010).
  \bibitem {ELS2004} S. Eisebitt, J. L\"uning, W. F. Schlotter, M. L\"orgen, O. Hellwig, W. Eberhardt, and J. St\"ohr, ''Lensless imaging of magnetic nanostructures by X-ray spectro-holography,'' Nature \textbf{432}, 885-888 (2004).
  \bibitem{C2011} H. N. Chapman, P. Fromme, A. Barty, T. A. White, R. A. Kirian, A. Aquila, M. S. Hunter, J. Schulz, D. P. DePonte, U. Weierstall, R. B. Doak, F. R. N. C. Maia, A. V. Martin, I. Schlichting, L. Lomb, N. Coppola, R. L. Shoeman, S. W. Epp, R. Hartmann, D. Rolles, A. Rudenko, L. Foucar, N. Kimmel, G. Weidenspointner, P. Holl, M. Liang, M. Barthelmess, C. Caleman, S. Boutet, M. J. Bogan, J. Krzywinski, C. Bostedt, S. Bajt, L. Gumprecht, B. Rudek, B. Erk, C. Schmidt, A. H\"omke, C. Reich, D. Pietschner, L. Str\"uder, G. Hauser, H. Gorke, J. Ullrich, S. Herrmann, G. Schaller, F. Schopper, H. Soltau, K. U. K\"uhnel, M. Messerschmidt, J. D. Bozek, S. P. Hau-Riege, M. Frank, C. Y. Hampton, R. G. Sierra, D. Starodub, G. J. Williams, J. Hajdu, N. Timneanu, M. M. Seibert, J. Andreasson, A. Rocker, O. J\"onsson, M. Svenda, S. Stern, K. Nass, R. Andritschke, C. D. Schr\"oter, F. Krasniqi, M. Bott, K. E. Schmidt, X. Wang, I. Grotjohann, J. M. Holton, T. R. M. Barends, R. Neutze, S. Marchesini, R. Fromme, S. Schorb, D. Rupp, M. Adolph, T. Gorkhover, I. Andersson, H. Hirsemann, G. Potdevin, H. Graafsma, B. Nilsson, and J. C. H. Spence, ''Femtosecond X-ray protein nanocrystallography,'' Nature \textbf{470}, 73-77 (2011).
  \bibitem{SEM2011} M. Seibert, T. Ekeberg, F. R. N. C. Maia, M. Svenda, J. Andreasson, O. J\"onsson, D. Odi\`c, B. Iwan, A. Rocker, D. Westphal, M. Hantke, D. P. DePonte, A. Barty, J. Schulz, L. Gumprecht, N. Coppola, A. Aquila, M. Liang, T. A. White, A. Martin, C. Caleman, S. Stern, C. Abergel, V. Seltzer, J. Claverie, C. Bostedt, J. D. Bozek, S. Boutet, A. A. Miahnahri, M. Messerschmidt, J. Krzywinski, G. Williams, K. O. Hodgson, M. J. Bogan, C. Y. Hampton, R. G. Sierra, D. Starodub, I. Andersson, S. Bajt, M. Barthelmess, J. C. H. Spence, P. Fromme, U. Weierstall, R. Kirian, M. Hunter, R. B. Doak, S. Marchesini, S. P. Hau-Riege, M. Frank, R. L. Shoeman, L. Lomb, S. W. Epp, R. Hartmann, D. Rolles, A. Rudenko, C. Schmidt, L. Foucar, N. Kimmel, P. Holl, B. Rudek, B. Erk, A. H\"omke, C. Reich, D. Pietschner, G. Weidenspointner, L. Str\"uder, G. Hauser, H. Gorke, J. Ullrich, I. Schlichting, S. Herrmann, G. Schaller, F. Schopper, H. Soltau, K. K\"uhnel, R. Andritschke, C. Schr\"oter, F. Krasniqi, M. Bott, S. Schorb, D. Rupp, M. Adolph, T. Gorkhover, H. Hirsemann, G. Potdevin, H. Graafsma, B. Nilsson, H. N. Chapman, and J. Hajdu, ''Single mimivirus particles intercepted and imaged with an X-ray laser,'' Nature \textbf{470}, 78-81 (2011).
  \bibitem{MYV2010} A. P. Mancuso, O. M. Yefanov, I. A. Vartanyants,
                    ''Coherent diffractive imaging of biological samples at synchrotron and free electron laser facilities,'' J. of Biotechnology, \textbf{149}, 229 (2010).
  \bibitem{V2007} I. A. Vartanyants, I. K. Robinson, I. McNulty, C. David, P. Wochner, and T. Tschentscher, ''Coherent X-ray scattering and lensless imaging at the European XFEL Facility,'' J. Synchrotron Radiat. \textbf{14}, 453-470 (2007).
  \bibitem{Y2011} L. Young, E. P. Kanter, B. Kr\"assig, Y. Li, A. M. March, S. T. Pratt, R. Santra, S. H. Southworth, N. Rohringer, L. F. Dimauro, G. Doumy, C. A. Roedig, N. Berrah, L. Fang, M. Hoener, P. H. Bucksbaum, J. P. Cryan, S. Ghimire, J. M. Glownia, D. A. Reis, J. D. Bozek, C. Bostedt, and M. Messerschmidt, ''Femtosecond electronic response of atoms to ultra-intense X-rays,'' Nature \textbf{466}, 56-61 (2010).
  \bibitem{WWQ2009} L. W. Whitehead, G. J. Williams, H. M. Quiney, D. J. Vine, R. A. Dilanian, S. Flewett, K. A. Nugent, A. G. Peele, E. Balaur, and I. McNulty, ''Diffractive imaging using partially coherent x rays,'' Phys. Rev. Lett. \textbf{103}, 243902 (2009).
  \bibitem{AWQ2011} B. Abbey, L. W. Whitehead, H. M. Quiney, D. J. Vine, G. A. Cadenazzi, C. A. Henderson, K. A. Nugent, E. Balaur, C. T. Putkunz, A. G. Peele, G. J. Williams, and I. McNulty, ''Lensless imaging using broadband X-ray sources,'' Nat. Photonics  \textbf{5}, 420 (2011).
  \bibitem{JY2010} Y. H. Jiang, T. Pfeifer, A. Rudenko, O. Herrwerth, L. Foucar, M. Kurka, K. U. K\"uhnel, M. Lezius, M. F. Kling, X. Liu, K. Ueda, S. D\"usterer, R. Treusch, C. D. Schr\"oter, R. Moshammer, and J. Ullrich, ''Temporal coherence effects in multiple ionization of N2 via XUV pump-probe autocorrelation,'' Phys. Rev. A \textbf{82}, 041403 (2010).
 \bibitem{FLASH} http://hasylab.desy.de/facilities/flash/publications/selected\_publications/index\_eng.html
 \bibitem{SBF2007} A. A. Sorokin, S. V. Bobashev, T. Feigl, K. Tiedtke, H. Wabnitz, and M. Richter, "Photoelectric effect at ultrahigh intensities," Phys. Rev. Lett. \textbf{99}, 213002 (2007).
 \bibitem{HFF2010} K. Honkavaara, B. Faatz, J. Feldhaus, S. Schreiber, R.Treusch, J.Rossbach,
                    ''FLASH UPGRADE'',
                    First International Particle Accelerator Conference, IPAC'10, Kyoto, Japan, (2010).
%
%
\bibitem{SVK2008} A. Singer, I. A. Vartanyants, M. Kuhlmann, S. D\"usterer, R. Treusch, and J. Feldhaus, ''Transverse-coherence properties of the free-electron-laser FLASH at DESY,'' Phys. Rev. Lett. \textbf{101}, 254801 (2008).
  \bibitem{VMS2010} I. A. Vartanyants, A. P. Mancuso, A. Singer, O. M.  Yefanov, and J. Gulden,
                    ''Coherence measurements and coherent diffractive imaging at FLASH,''
                    J. Phys. B \textbf{43}, 194016 (2010).
  \bibitem{RSW2011} S. Roling, B. Siemer, M. W\"ostmann, H. Zacharias, R. Mitzner, A. Singer, K. Tiedtke, and I. A. Vartanyants, ''Temporal and spatial coherence properties of free-electron-laser pulses in the extreme ultraviolet regime,''
                    Phys. Rev. ST Accel. Beams \textbf{14}, 080701 (2011).
  \bibitem{SE2010} E. L. Saldin, E. A. Schneidmiller and M. V. Yurkov,
                    \textit{The Physics of Free Electron Lasers}, (Springer, 2010).
  \bibitem{VSM2011} I. A. Vartanyants,
     A. Singer, A. P. Mancuso, , O. M. Yefanov, A. Sakdinawat, Y. Liu, E. Bang, G. J. Williams, G. Cadenazzi, B. Abbey, H. Sinn, D. Attwood, K. A. Nugent, E. Weckert, T. Wang, D. Zhu, B. Wu, C. Graves, A. Scherz, J. J. Turner, W. F. Schlotter, M. Messerschmidt, J. L\"uning, Y. Acremann, P. Heimann, D. C. Mancini, V. Joshi, J. Krzywinski, R. Soufli, M. Fernandez-Perea, S. Hau-Riege, A. G. Peele, Y. Feng, O. Krupin, S. Moeller, and W. Wurth, ''Coherence Properties of Individual Femtosecond Pulses of an X-Ray Free-Electron Laser,'' Phys. Rev. Lett. \textbf{107}, 144801 (2011).
  \bibitem{G2000} J. W. Goodman, \textit{Statistical Optics} (Wiley, New York, 2000).
  \bibitem{MW1995} L. Mandel and E. Wolf, \textit{Optical Coherence and Quantum Optics} (Cambridge University Press, 1995).
  \bibitem{SW2010} W. F. Schlotter, F. Sorgenfrei, T. Beeck, M. Beye, S. Gieschen, H. Meyer, M. Nagasono, A. F\"ohlisch, and W. Wurth, ''Longitudinal coherence measurements of an extreme-ultraviolet free-electron laser'',
                    Optics Letters \textbf{35}, 372-374 (2010).
  \bibitem{SF2010} F. Sorgenfrei, W. F. Schlotter, T. Beeck, M. Nagasono, S. Gieschen, H. Meyer, A. F\"ohlisch, M. Beye, and
                    W. Wurth, "The extreme ultraviolet split and femtosecond delay unit at the plane grating monochromator beamline PG2 at FLASH",
                    Rev. Sci Instr. \textbf{81}, 043107 (2010).
  \bibitem{MR2009} R. Mitzner, B. Siemer, M. Neeb, T. Noll, F. Siewert, S. Roling, M. Rutkowski, A. A. Sorokin, M. Richter, P. Juranic, K. Tiedtke, J. Feldhaus, W. Eberhardt, and H. Zacharias,
                    ''Spatio-temporal coherence of free electron laser pulses in the soft x-ray regime'',
                    Optics Express \textbf{16}, 19909-19919 (2009).
  \bibitem{SSY2008} E. L. Saldin, E. A. Schneidmiller, and M. V. Yurkov,
                    ''Coherence properties of the radiation from X-ray free electron laser'',
                    Opt. Communications \textbf{281}, 1179-1188 (2008).
  \bibitem{VS2010} I. A. Vartanyants and A. Singer,
                    ``Coherence properties of hard x-ray synchrotron sources and x-ray free-electron lasers,''
                    New J. Phys. \textbf{12}, 035004 (2010).
  \bibitem{BPG2002} R. A. Bartels,  A. Paul, H. Green, H. C. Kapteyn, M. M. Murnane, S. Backus, I. P. Christov, Y. Liu,
                    D. Attwood, and C. Jacobsen,
                    ''Generation of Spatially Coherent Light at Extreme Ultraviolet Wavelengths,''
                    Science, \textbf{297}, 376-378 (2002).
  \bibitem{MB2010} B. W. J. McNeil and N. R. Thompson,
                    " X-ray free-electron lasers",
                    Nat. Photon., \textbf{12}, 814-821, (2010).
  \bibitem{S2009} F. Staier,
                    ''Entwicklung, Bau und Test einer UHV R\"ontgenstreukammer f\"ur die digitale In-Line Holographie,''
                    PhD Thesis (University of Heidelberg, 2009).
                  \bibitem{WM2007}  M. Wellh\"ofer, M. Martins, W. Wurth, A. Sorokin, and M. Richter, ''Performance of the monochromator beamline at FLASH,'' J. Opt. A, Pure Appl. Opt. \textbf{9}, 749-756 (2007).
  \bibitem{footnote1} The assumptions used in deriving equation (\ref{eq:fit}) were well satisfied in our
                        experimental geometry. The maximum time delay introduced through the path length difference
                        was $\tau_{max}\approx0.6$ fs and was smaller than the temporal coherence length $\tau_c=(1.75\pm0.0.01)$ fs measured by the split and delay unit (see below). Therefore, we could safely assume that in transverse coherence measurements $|\gamma_{12}^{\rm eff}(\tau)|\approx|\gamma_{12}^{\rm eff}(0)|$ and $\alpha_{12}(\tau)\approx\alpha_{12}(0)$.
  \bibitem{footnote4} An unconstrained fit yields a value of $|\gamma_{11}^{\rm eff}|\approx0.8$ in both directions and
                    provides slightly larger values for the transverse coherence length. We attribute this to inhomogenities in the transmission through the pinholes.
 \bibitem{CKJ2010} J. Chalupsky, J. Krzywinski, L. Juha, V. Hajkova, J. Cihelka, T. Burian, L. Vyain, J. Gaudin, A. Gleeson, M. Jurek, A. R. Khorsand, D. Klinger, H. Wabnitz, R. Sobierajski, M. St\"ormer, K. Tiedtke, and S. Toleikis,
   ''Spot size characterization of focused non-Gaussian X-ray laser beams'',
   Opt. Express \textbf{18}, 27836 (2010).
  \bibitem{footnote5} We attribute this positional uncertainty to both, instabilities
    of the sample stages and beam positional jitter.
%
%
  \bibitem{footnote3} In our experiment the maximum of $|\gamma_{12}(\tau)|$ did not reach unity but rather a value
                    of $0.14$. The reason for this is that the full beam was split in the middle and overlapped again meaning that parts of the center of the beam were overlapped with parts of the edge of the beam (see \cite{SF2010} for details). This corresponds to a large pinhole separation in a Young's double pinhole experiment yielding reduced values of $|\gamma_{12}(0)|$. The beam was not spatially filtered with the apertures of the PG2 beamline, which would increase the contrast. Additionally, Ce:YAG crystals are known to saturate at high intensities \cite{BAS2009}, which can result in a degradation of fringe visibility.
 \bibitem{BAS2009} D.P.~Bernstein, Y.~Acremann, A.~Scherz, M.~Burkhardt, J.~Stöhr, M.~Beye, W. F.~Schlotter, T.~Beeck,
                    F.~Sorgenfrei,  A.~Pietzsch, W.~Wurth, A. F\"{ö}hlisch,
                    ''Near edge x-ray absorption fine structure spectroscopy with x-ray free-electron lasers,''
                    Appl. Phys. Lett. \textbf{95}, 134102 (2009).
  \bibitem{parameters} We used the following FLASH operation parameters \cite{FLASH2012,Faatz2012} $K=1.23$, $\lambda_w=27.3$
                    mm, $I=2200\pm300$ A, $\gamma=1741$, $\sigma_{\bot}=95\pm35~\mu$m, and $A_{JJ}=0.83$ to calculate the FEL parameter $\rho$ in equation (\ref{eq:coherencetime}). The error is derived by applying the Gaussian error propagation law.
  \bibitem{FLASH2012} http://flash.desy.de/accelerator/, access December 2011.
  \bibitem{Faatz2012} B. Faatz, private communication.
\end{thebibliography}
\end{document}